\documentclass{article}
\usepackage{arxiv}

\usepackage{booktabs} 
\usepackage{multirow} 
\usepackage{siunitx} 
\usepackage{amssymb} 
\usepackage{soul,xcolor} 
\usepackage{comment} 
\usepackage{listings} 
\usepackage{fancyvrb} 
\usepackage{hyperref} 
\usepackage{graphicx} 

\begin{document}
\title{Sharing pandemic vaccination certificates through blockchain: Case study and performance evaluation}

\author{
 Jos\'{e} L. Hern\'{a}ndez-Ramos, Georgios Karopoulos, Dimitris Geneiatakis, Tania Martin,\\
 \textbf{Georgios Kambourakis, and Igor Nai Fovino} \\
  European Commission\\
  Joint Research Centre\\
  Ispra 21027, Italy \\
  \texttt{\{jose-luis.hernandez-ramos, georgios.karopoulos, dimitrios.geneiatakis,}\\ \texttt{tania.martin, georgios.kampourakis, igor.nai-fovino\}@ec.europa.eu}
}
\thanks{This work has been submitted to the IEEE for possible publication. Copyright may be transferred without notice, after which this version may no longer be accessible.}

\maketitle

\begin{abstract}
This work proposes a scalable, blockchain-based platform for the secure sharing of COVID-19 or other disease vaccination certificates. As an indicative use case, we simulate a large-scale deployment by considering the countries of the European Union. The proposed platform is evaluated through extensive simulations in terms of computing resource usage, network response time and bandwidth. Based on the results, the proposed scheme shows satisfactory performance across all major evaluation criteria, suggesting that it can set the pace for real implementations. Vis-\`a-vis the related work, the proposed platform is novel, especially through the prism of a large-scale, full-fledged implementation and its assessment.
\end{abstract}

\section{Introduction}\label{sec:introduction}

To defend against the COVID-19 pandemic, several initiatives and actions have been hitherto undertaken, including rapid diagnosis and isolation of infected people, as well as the creation of digital contact tracing frameworks \cite{martin2020demystifying}. However, the second COVID-19 wave during the fall of 2020 showed that these measures are insufficient, especially when they are abruptly relaxed. Naturally, in the mid or long term, the enforcement of such measures is sure to induce serious socioeconomic consequences. Therefore, the sheer objective has been the development of effective and safe vaccines to be rolled out worldwide. Indeed, numerous efforts have been initiated during 2020 involving medical institutions, pharmaceutical companies and research centers worldwide to get a vaccine at unprecedented speed. In Nov. 2020~\cite{who_vaccines2020}, there were 48 and 164 vaccines in clinical and preclinical evaluation, respectively.

While the realisation of vaccines represents the main objective to terminate the pandemic, their
deployment 
is also associated with important challenges. 
The vaccination process will be prioritised for certain population groups according to different aspects, such as age, health condition, and profession. Hence, immune and vulnerable people will live together during a certain period of time. In this situation, the use of digital vaccination certificates could help alleviate the burden on health systems, as vaccinated people would not need to perform viral tests, which are currently required to, say, travel to different countries. Unlike the current paper version of vaccination certificates, namely the International Certificate of Vaccination or Prophylaxis (ICVP), these digital documents would allow a far more scalable solution along with a faster and more secure verification process \cite{harvard_2020}.

Blockchain technology has been already identified as a promising approach to combat the pandemic in different scenarios, such as early detection of outbreaks, medical supply chain, or donation tracking \cite{kalla2020role}. In the same mindset, the creation of a blockchain platform to share information about the pandemic would increase transparency, interoperability and accountability so that potential discrepancies among data from different sources, say, medical centers or governments, could be avoided. This would foster a more trustworthy reporting and monitoring of the pandemic evolution considering diverse territories and countries. Furthermore, such a platform would increase citizens’ trust in the vaccination process, as the information related to vaccines could be publicly available \cite{lazarus2020global}.

This work analyses the main requirements to build a scalable platform for sharing vaccination data, and the advantages of blockchain for the realisation of such a platform. We focus on the scenario of vaccination certificates that can be generated after a citizen is vaccinated, and how blockchain could aid in maintaining such information towards enabling a secure and privacy-aware verification process. 
Furthermore, we provide a comprehensive performance evaluation of the proposed platform by considering the vaccination of the EU population and 27 blockchain nodes, representing each Member State (MS) in the EU. We meticulously assess our platform under different network conditions, including latency and bandwidth. To our knowledge, this is the first work to provide an estimation of the performance requirements associated to a blockchain-based platform for vaccination data in a large scale. Furthermore, we discuss practical aspects and security considerations for a large-scale deployment of the intended platform, along with potential regulatory implications of vaccination certificates.

\section{Related work}\label{sec:related}

From the existing works, three different types of digital health certificates can be identified: (a) \emph{vaccination certificates}, referring to whether a person has received the vaccine or not, (b) \emph{diagnostic test certificates}, demonstrating that a person has undergone a test, and (c) \emph{immunity certificates} or \emph{immunity passports}, implying that a person has developed antibodies.

Two works address both vaccination and immunity certificates simultaneously.
The only peer-reviewed one is~\cite{Eisenstadt2020}, which is based on Verifiable Credentials (VC) as digital IDs, the decentralised data storage platform Solid, and a consortium Ethereum-based blockchain. 
The other work is~\cite{hasan_blockchain-based_2020}, which is based on Ethereum smart contracts, Self-Sovereign Identity (SSI), and InterPlanetary File System (IPFS) to store medical tests and travel history in a decentralised manner.

Regarding vaccination certificates only, the authors of~\cite{Chaudhari2020} focus on privacy and propose a hashing algorithm that enables users store the information on the blockchain anonymously using an ID that is created from their iris.
In this case, the vaccination certificate data and a hash of the user ID are stored on the blockchain. This could imply a potential issue since it would demand a very high storage requirements of the blockchain nodes. This is especially true in the case of populous or multiple countries using the same blockchain.

The authors of~\cite{Angelopoulos2020} introduce the concept of digital health passports, which is similar to the diagnostic test results required for travellers in certain cases.
It is based on a private blockchain using the proof-of-authority consensus mechanism, where the test results are registered and stored.

For immunity certificates, the work of~\cite{Hicks2020} presents SecureABC, a privacy-oriented protocol based on public key cryptography.
This proposal does not use blockchain and the certificates can be either paper or app-based. As a consequence, if the paper certificate or the mobile device are lost, so are the respective certificates.
In~\cite{Bansal2020}, the concept of COVID-19 immunity certificates is using a government-run blockchain.

None of all these works presents concrete experimental results with technical details or proof-of-concept implementation. They are generally incomplete. For instance, they rather provide simple short high-level descriptions of the proposed solutions, or unconvincing benchmarks, limited to a number of simultaneous requests, thus being far from real-world deployment scenarios. In contrast, our work tackles the problem from an EU-27 perspective.
Nonetheless, most of the works analyse the anonymity and privacy concerns regarding vaccinated/tested persons.

Our work concentrates on vaccination certificates, influenced by the views of WHO on immunity passports: \textit{``...there is not enough evidence about the effectiveness of antibody-mediated immunity to guarantee the accuracy of an `immunity passport'.''}
Another aspect driving us to this direction is that vaccination certificates will incite people to get vaccinated, while immunity certificates could motivate individuals get infected for possessing the necessary antibodies.

\section{The need for a data-driven platform}\label{sec:covid-blockchain}

The global deployment of COVID-19 vaccines sets out unprecedented challenges to be addressed in the period ahead, including an efficient supply chain and effective monitoring of vaccination coverage in a certain region. Indeed, in the case of two-shot vaccines, more than 15 billion vaccines would be required to be distributed and deployed worldwide.

At the European level, the Commission published in Oct. 2020 a document on COVID-19 vaccination strategies and vaccine deployment for the 27 MS~\cite{ec2020}. This document establishes the need to define a common strategy for the vaccination process, promoting coordination and collaboration among EU countries. One of the main goals of this strategy is to increase the acceptance of COVID-19 vaccines. Actually, recent studies reveal that a significant part of the population would not be willing to be vaccinated against the COVID-19 disease~\cite{lazarus2020global}. To address this issue, there is a need for an effective, consistent, and transparent communication of information related to COVID-19 vaccines and the vaccination process itself. As described in~\cite{ec2020}, the sharing of pandemic related information among MS would cater for a better monitoring of the different vaccines under development, including data on possible side effects, which would be made readily available to the relevant authorities.

For the realisation of this COVID-19 data sharing platform, blockchain technology has been postulated in different related scenarios, including contact tracing and outbreaks, where information sharing is essential. Blockchain is based on a distributed ledger that is shared by a set of entities. The ledger contains a list of immutable transactions that are validated by the participating entities through a consensus mechanism.
Furthermore, a blockchain can be permissionless (any entity can participate) or permissioned (participation is limited to a set of entities). The development of a blockchain-based platform offers a high degree of transparency and accountability, fostering a trustworthy environment for the sharing of COVID-19 data.

While a blockchain platform for sharing pandemic data could help in different scenarios, we focus on the use of blockchain to register the data associated to a vaccinated citizen. The envisioned platform will enable a trusted ecosystem to track the deployment of vaccines in a certain region and consider priority groups. For this purpose, we examine the concept of digital vaccination certificates that could be demonstrated by citizens to carry out certain activities without the need of PCR tests. The following sections describe the design and architecture for the use of digital vaccination certificates, as well as a thorough evaluation considering a blockchain platform where each MS is represented by a blockchain node.

\section{Designing vaccination certificates}\label{sec:idm}

As mentioned in Section~\ref{sec:related}, the implementation of COVID-19 related certificates attracted an increasing interest. However, since immunity certificates have received criticism from different sources, this work concentrates on vaccination certificates associated with an individual who has been vaccinated. Indeed, the WHO has established a recent initiative with the Estonian government to create a blockchain-based vaccination certificate infrastructure. These certificates can be viewed as a digital version of the ICVP certificates created by WHO that show a person's vaccines and the date they were provided. Our work is aligned with this initiative by developing a blockchain platform to analyse the performance requirements of exploiting these certificates.

For the representation of this certificate, there is a need to create an appropriate identity management (IdM) ecosystem to link the vaccination of a citizen with their identity, so that the resulting certificate is verifiable, scalable, and privacy-preserving. In recent years, the concept of Self-Sovereign Identity (SSI) has emerged as a decentralised alternative to traditional centralised IdM systems in which an identity provider has control over the user's identity data for the sake of providing authentication and registration services. In an SSI system, the lack of a central entity allows users a selective disclosure of their identity information as they are empowered to use different identities depending on the transaction being performed. 
Since it was coined in 2016, SSI has been linked to the use of Decentralised Identifiers (DID) and Verifiable Credentials (VC) that are being standardised by the World Wide Web Consortium (W3C). A DID is an identifier under the control of a DID subject that indicates a DID method and a specific identifier of such method. Furthermore, a DID is resolved to a DID document that specifies verification methods for that DID. 
While DIDs foster a decentralised authentication process, they do not provide information about the subject itself that may be required to perform certain online operations, say, being over 18. For this reason, DIDs are usually used together with VCs, which represent a digital version of a paper certificate in which a certain entity (issuer) asserts certain information (claims) about a subject in a way that can be verified by other entities (verifiers). The VCs data model is a W3C recommendation that allows to demonstrate certain identity information through a verifiable presentation. Additionally, it may allow a selective disclosure of identity data through the use of zero-knowledge proofs to preserve users' privacy. 

The use of VCs in the context of the COVID-19 crisis has been fostered by the COVID-19 Credentials Initiative, which groups around 100 organisations to support efforts of using VCs to mitigate the spread of the virus. While it is not the focus of our work, based on ongoing discussions about vaccination certificates, we believe that the following aspects should be considered when creating such credentials:

\begin{itemize}

    \item The format of the VC representing vaccination certificates should be publicly available, and a URI to such scheme should be included in the \textit{context} field of the VC.
    \item The issuer of the VC could be delineated by the medical center itself or the physician who administered the vaccine to the citizen. This entity could be identified through a DID in the VC that can be registered in the blockchain for validation purposes. 
    \item The signature of the VC, i.e., the \textit{proof} field, should be based on standard formats, such as JSON Web Signature.
    \item The VC \textit{subject} is represented by the citizen getting vaccinated, and could also be identified by a DID linked to the identity proven during the vaccination process. 
    \item Some of the vaccines under development require two shots to be effective. Therefore, vaccination certificates could be only issued when both shots are administered to the citizen. Alternatively, each vaccination shot could be associated to a VC so that it can be used to prove that the citizen already received the first shot. 
    \item The validity of the validation certificate, namely the \textit{expirationDate} field, should be associated with the period during which this vaccination is effective. This time could differ among the vaccines under development. 
    \item The VC should contain additional information about the vaccine being administered, as well as data about the specific doses of a certain vaccine. These pieces of information could be used to track the coverage and deployment of different vaccines and may aid in the management of the supply chain.
\end{itemize}

Recall that the design of a VC for vaccination certificate is out of the scope of this paper. Indeed, only a hash of such credential will be stored in the blockchain platform, which is described in the next section. 

\section{Vaccination certificate scenario}\label{sec:scenario}

\begin{figure}[htp]
    \centering
    \includegraphics[width=10cm]{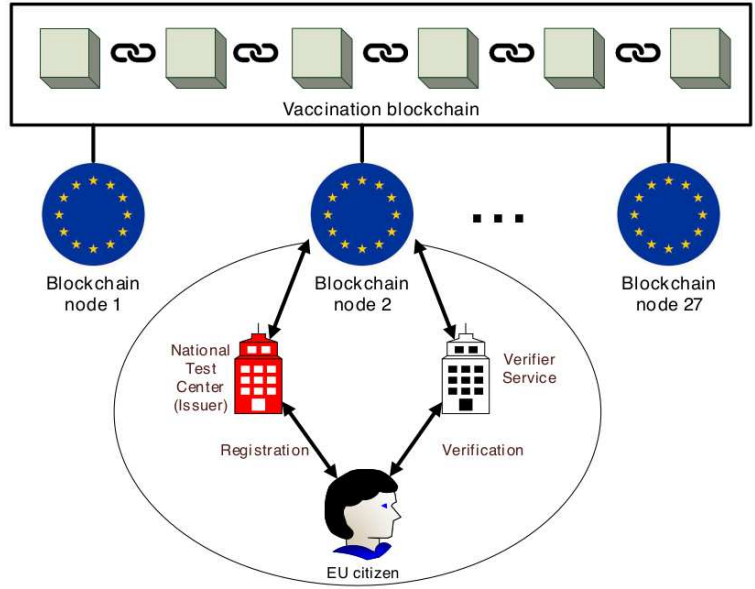}
    \caption{Overview of the proposed blockchain-based vaccination certificate platform}
    \label{fig:scenario}
\end{figure}

For the development of the proposed blockchain platform, we consider the architecture proposed in Figure \ref{fig:scenario}. Naturally, the depicted architecture does not reflect the reality of any decision made at EU level, but it solely serves as a proof-of-concept for evaluation purposes. The architecture includes the \textit{vaccination blockchain}, which represents a permissioned blockchain to store vaccination information related to EU countries. Each of the 27 MSs can designate a blockchain client/node to interact with the blockchain that can be represented by a national health authority, say, the ministry of health. This entity is also in charge of designating a set of national medical centers to generate vaccination certificates associated to already vaccinated people. These certificates will be validated by \textit{verification centers}, which represent any organisation, public or private (e.g., airport or public administration building), that needs to verify the vaccination status of an individual.

The vaccination blockchain is used to store all relevant information about the vaccination process, including the registration of national medical centers, say, name and address. The registration of these entities can be performed by the national health authorities, which represent the blockchain nodes of their MS, through the use of smart contracts. Furthermore, the blockchain will simply contain a hash of the vaccination certificate per citizen that will be generated during the registration process, and used later to facilitate the process of verifying the vaccination status of a citizen. It is important to note that the registration and verification processes analysed below are only illustrative examples of how the proposed platform can be used to manage vaccination certificates, and they are considered for evaluation purposes in the next section. Furthermore, while both aforementioned processes are described based on the use of VCs, other formats for vaccination certificates could be also used in our platform.

During the \textit{registration} process, citizens go to a national medical center, where they present a valid identity document, to get vaccinated. Then, a physician performs the vaccination and the corresponding vaccination certificate is generated. As described in Section \ref{sec:idm}, this certificate may contain information about the vaccine itself, as well as data about the specific dose to facilitate the management of the supply chain. Furthermore, the citizen's identity shown at the beginning of the process can be used to generate a DID that is embedded in the credential. As discussed in Section \ref{sec:idm}, in the case of a two-shot vaccine, this credential may represent that the citizen received the first shot, so it can be used in the process of administering the second one, or that the person is immune as they already received both shots. The national medical center, or the physician on its behalf, signs the VC to guarantee its validity and sends this credential to the citizens, say, through an app in their smartphone, so that they can maintain the control of how the VC is used. Furthermore, a hash of the VC is generated and stored on the blockchain. The national medical center sends this hash to the MS's blockchain node that is responsible for registering it on the EU vaccination blockchain. For this process, as described in the next section, we consider that the consensus mechanism is based on the approach followed by \emph{Hyperledger}. Additionally, an encrypted version can be stored in IPFS or another repository so that the vaccination certificate can be recovered by the citizen in case of losing the smartphone.

After citizens have received a vaccination certificate based on VC, they can use it to access certain places that require proof of citizens' vaccination status, such as an airport or public administration building. The \textit{verification} process of the VC is carried out through a ``verifiable presentation'', in which citizens present their vaccination certificate to a verifier service. This service creates a hash of the provided certificate that is verified against the hash stored in the blockchain. For this process, the national medical center contacts the country's blockchain node that is in charge of performing the verification on the EU vaccination blockchain. Furthermore, the verification service validates the signature created by the national medical center to confirm the credential was generated by an approved entity. In addition, it checks the validity of the VC, as well as the citizen to ensure that it is indeed the person associated with the credential presented. These checks can be performed by using the DIDs associated to the issuer (i.e., the national medical center) and subject (i.e., the citizen) that were embedded in the VC. Alternatively, the user can carry out a verifiable presentation by showing a subset of the attributes contained in the credential through the use of zero knowledge proofs. Actually, this process can be useful to access certain places that only require confirmation of a person's vaccination status, but do not need personal data. However, although the use of zero knowledge proofs is contemplated by the VC specification, it is outside the scope of this work. Indeed, as described in the next section, the evaluation of our platform is focused on the performance requirements from the perspective of the blockchain implementation to register and verify vaccination certificates.

\section{Evaluation}\label{sec:evaluation}

\subsection{Testbed}\label{sec:testbed}

To evaluate our proposal, we rely on the Experimental Platform for Internet Contingencies (EPIC)~\cite{6646193}.
EPIC is an emulation testbed based on the Deter software~\cite{
Benzel:2011:SCS:2076732.2076752} for studying the security and stability of distributed systems.
The infrastructure of EPIC comprises 356 experimental nodes, 8 switches and a few special equipment, such as Programmable Logical Controllers.

Overall, the setup relies on the deployment of Hyperledger Fabric on an emulated network in EPIC, and implements the proposed architecture shown in Figure~\ref{fig:scenario}. It is assumed that the European health authorities, which are considered trusted, provide the ``ordering'' services, while each MS is a ``peer'' node in the Hyperledger Fabric terminology.

The ordering services comprise the following: ZooKeeper (3 instances), Kafka (4 instances) and Orderer (3 instances). Their main purpose is to sort the messages/requests exchanged among the participants. Each instance of a given service runs on a different machine for supporting fail-over of the ordering services. This setup ensures ordering service availability if at maximum one instance of each service is in fail status. The peer nodes are managed by the MSs for endorsing the transactions proposed by the clients. They also receive the ordered blocks of transactions from the ordering service to maintain their local copy of the ledger. The following services of a MS node are hosted on a single machine:

\begin{enumerate}
 \item \textit{CouchDB:} A database that maintains the valid transactions of the blockchain and allows content-based JSON queries. 
 \item \textit{Peer:} A core service in the Hyperledger Fabric architecture  storing the ledger and validating the transactions.
\item \textit{Certificate Authority:} Provides digital certificates to the participants of the MS node. 
\item \textit{Smart contract:} Implements basic functionalities such as 
user access control and message conformity. 
\item \textit{Application interface:} Interacts with the blockchain. It is implemented as a representational state transfer (REST) service and accomplishes all the interactions on behalf of the national health centers for committing a transaction in the blockchain network. 
\end{enumerate}

All ordering and peer services are configured and executed using the corresponding docker images with the standard deployment options. 
Moreover, all the underlying network communications among the participants (clients, peers, and the ordering service), are securely protected by TLS. 
The certificates and private keys both for TLS and the blockchain services are generated during the blockchain network initialisation procedure, according to the Hyperledger Fabric specifications.

\subsection{Results}

We evaluate the adequacy of deploying our proposal in a real, large-scale architecture, concentrating on two fundamental provisioned services, namely vaccination registration and verification. Precisely, the emulated 1~Gbps blockchain network comprises 27~nodes corresponding to the current EU member states with a network latency of 3~msec.
As each MS acts independently, we deploy a single disjunctive (``OR'') policy among the participants, meaning that each of them is not subjective to a validation from the others, and the system checks whether the submitted transaction bears a valid digital signature from the MS blockchain node. This also means that any transaction stemming from another MS will be rejected by the blockchain. The focus is on user experience in terms of request round-trip time (i.e., the time required for receiving a response after submitting a request) and the utilisation of system resources (i.e., CPU, memory, and network bandwidth). 

For the registration process, we consider the maximum number of transactions required to get all European citizens vaccinated in one year. According to Eurostat, the EU-27 population is $\approx$447.5M inhabitants. Thus, assuming that a vaccine requires two doses, that is two blockchain transactions, a total of 28~transactions per second will be required in the worst case.
Table~\ref{tbl:net-eval} summarises the average latency perceived when registering or verifying a vaccination certificate in the blockchain, as well as the bandwidth consumed both by the peer and the ordering nodes. As observed, the response time for registration ranges between 83 and 133~msec.
Moreover, at the peer side, the bandwidth utilisation 
increases from 500 to 700~KB. Overall, both these numbers can be characterised as absolutely tolerable. On the other hand, the bandwidth consumed by the ordering service demonstrates a significant augmentation among the different TPS values, reaching $\approx$6000~KB in the most demanding case. 

\begin{table}
\caption{Network evaluation of registering and verifying vaccination certificates using blockchain}
\centering
{\begin{tabular}{lS[table-number-alignment = center]S[table-number-alignment = center]S[table-number-alignment = center]S[table-number-alignment = center]}
\toprule
{\multirow{2}{*}{\textbf{Step}}} & {\multirow{2}{*}{\textbf{TPS}}} & \textbf{Response} & \textbf{Peer} & \textbf{Ordering}\\
&  & \textbf{time (msec)} & \textbf{bandwidth (KB)} & \textbf{bandwidth (KB)}\\
\midrule
 \multirow{6}{*}{Register} &  1 & 84 & 395 & 636\\

  & 2 & 81 & 419 & 825\\

  & 4 & 78 & 457 & 2019\\

 & 8 & 87 & 516 & 3644\\

  & 16 & 109 & 588 & 4938\\

  & 28 & 133 & 700 & 6019\\

\midrule

 \multirow{8}{*}{Verify} &  1 & 91 & 394 & 701\\

  & 2 & 87 & 415 & 1123\\

  & 4 & 83 & 447 & 1788\\

 & 8 & 94 & 495 & 2153\\

  & 16 & 117 & 553 & 5069\\

  & 28 & 153 & 639 & 8122\\

  & 50 & 168 & 671 & 5919\\

  & 100 & 189 & 804 & 12109\\

\bottomrule
\end{tabular}}
\label{tbl:net-eval}
\end{table} 

\begin{figure}
    \centering
    \includegraphics{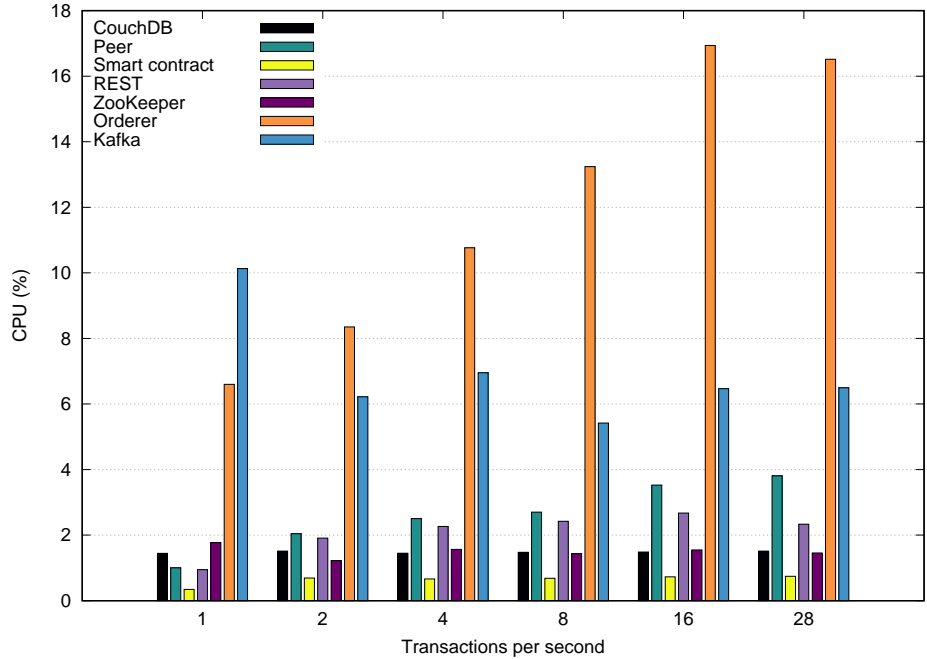}
    \caption{Dockerised services' CPU utilisation considering different TPS for registering new vaccination certificates in a blockchain system}
    \label{fig:cpuregistration}
\end{figure}

\begin{figure}
    \centering
    \includegraphics{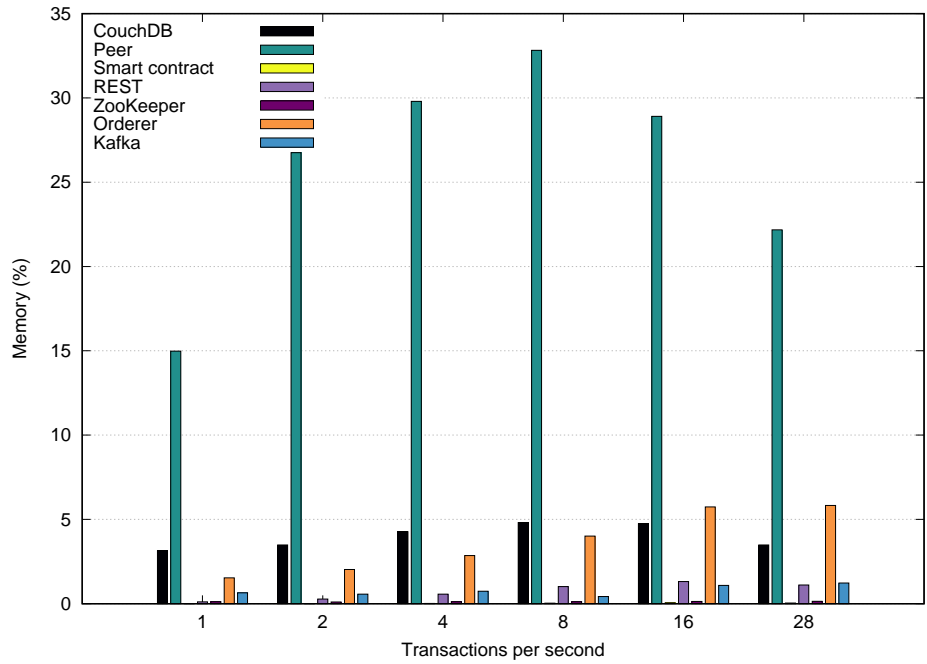}
    \caption{Dockerised services' memory utilisation considering different TPS for registering new vaccination certificates in a blockchain system}
    \label{fig:memregistration}
\end{figure}

CPU and memory utilisation under different traffic conditions per service are illustrated in Figures~\ref{fig:cpuregistration} and \ref{fig:memregistration}. Particularly, considering the worst case, CPU and memory utilisation for the peer services remain under 4 and 35\% respectively, while the ordering services utilisation is under 17 and 7\%.
In any case, these requirements for both services are manageable.
It is also perceived that, when TPS increase from 8 to 16 and above, memory utilisation for the peer services starts to decrease. This can be explained by the fact that, along with TPS, the response time augments, having transactions submitted to the system at a lower rate.
Interestingly and also on the positive side, CPU utilisation for the smart contract remains almost constant under different TPS, consuming less than 1\% of the available CPU cycles.
Overall, the registration process is more demanding in terms of CPU on the orderer and secondly on the kafka services, while in terms of memory on the peer service.

Regarding vaccination certificate verification,
we used data from Eurostat to calculate realistic requirements in terms of transactions per second.
We calculated the total number of air, marine, rail and bus passengers for 2018, which is the latest year with data for all these categories.
As verification transaction requests are forwarded to the national node of each MS, we consider the worst case, that is the MS with the highest combined number of passengers in one year (3.2~billion); this gives us $\approx$100~TPS.

Regarding the search operation, the worst case scenario is again followed, that is the correct record is the last one. Similar to vaccination registration, the response time increases proportionally to the number of TPS, demonstrating a similar pattern. Overall, with reference to Table~\ref{tbl:net-eval}, the response time and bandwidth utilisation at a MS blockchain node fluctuate between 91 and 189~msec and 394 and 804~KB respectively. However, for ordering, the utilised network bandwidth reaches up to 12109~KB.

\begin{figure}
    \centering
    \includegraphics{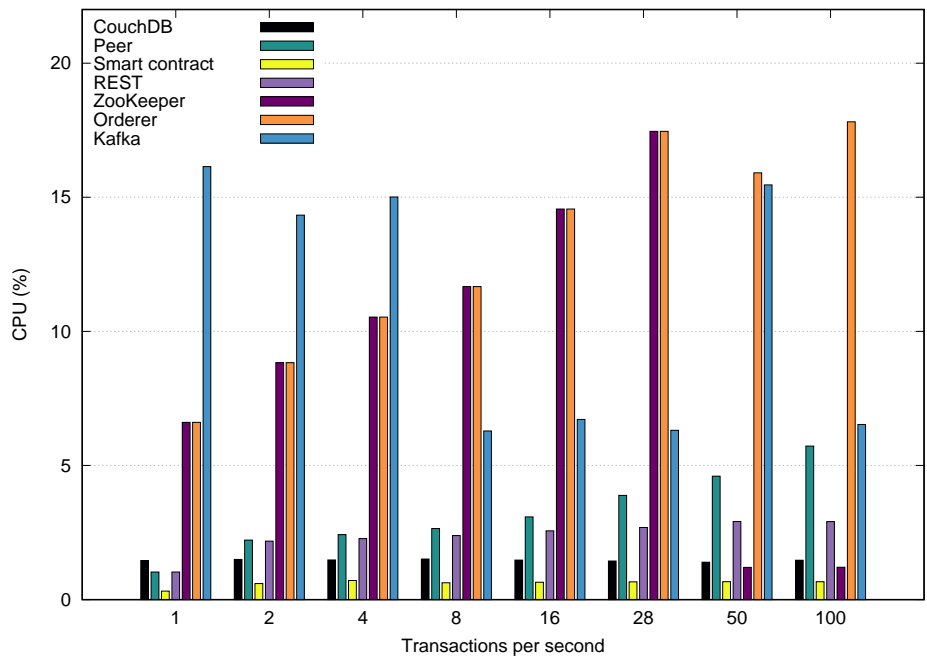}
    \caption{Dockerised services' CPU utilisation considering different TPS for verifying vaccination certificates in a blockchain system}
    \label{fig:cpuverification}
\end{figure}

\begin{figure}
    \centering
    \includegraphics{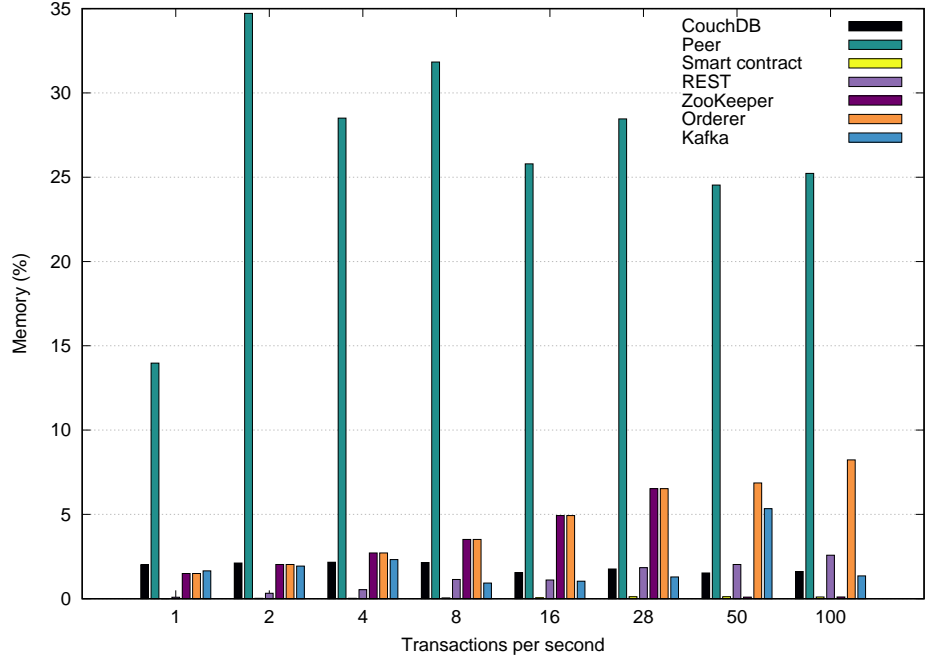}
    \caption{Dockerised services' memory utilisation considering different TPS for verifying vaccination certificates in a blockchain system}
    \label{fig:memverification}
\end{figure}

Figures \ref{fig:cpuverification} and \ref{fig:memverification} depict CPU and memory utilisation per blockchain service. As observed, CPU utilisation for both the peer and REST services increases proportionally to TPS, while it is relatively stable for couchDB and smart contract.
The orderer service initially increases and then stabilises, while the kafka service fluctuates between 6 and 16\%.
However, in all cases the CPU load remains under 18\%.
As expected, and similar to registration, memory usage for the peer service ranges between $\approx$14\% and 35\%, demonstrating that it is memory intensive. For the rest of the services, memory requirements are low, that is under 8\%.
In summary,  the  verification  process  is  more  demanding in  terms of  CPU  on  the  ordering services, while in terms of memory on the peer service.

\section{Conclusions}\label{sec:conclusion}

The work at hand sheds light on the timely issue of managing digital vaccination certificates on a large scale. After arguing that under the prism of COVID-19 and future epidemics this need is rather a sine qua non, we specifically attempt to answer two key questions: how such an endeavour can be realistically organised with a focus on reducing complexity, and if so, would it be smooth-running under pragmatic conditions or even stress in terms of performance? For the first matter, we scrutinised on an envisaged wide-scale deployment capable of covering the needs of EU-27, and elaborated on a practical vaccination certificate scenario. For the second, we relied on the EPIC platform.

Specifically, based on the performance results obtained, including scalability aspects and challenges for the deployment of such platform, it is demonstrated that, for both registration and verification operations, the system achieves satisfactory results even under stress. This strongly suggests that even a network decreased by one order of magnitude (100~Mbps) would be more than enough. Regarding CPU requirements, the ordering nodes need to be more powerful than MS ones, while the peer nodes necessitate more memory. Also, it is shown that, at least in a similar setup as our testbed, 100~TPS is the boundary, taking into account that above this limit the system is saturated, producing errors and experiencing inconsistencies. This indicates that in most populated European countries, the MS node specifications should be carefully devised to support such a large number of TPS or even greater if necessary.

Future work will concentrate more on the security, privacy, and ethical aspects associated with the registration and verification process of digital vaccination certificates. Also, an interesting direction is to investigate if this kind of platform could be also fruitful for serving the needs of the vaccines supply chain, ensuring efficient vaccine warehousing, handling, and stock administration.


\end{document}